# Two-component anomalous Hall and Nernst effects in anisotropic Fe$_{4-x}$Ge$_x$N thin films

Robin K. Paul [1], Jakub Vít [2], Petr Levinský [2], Jiří Hejtmánek [2], Ondřej Kaman [2], Mariia Pashchenko [2], Lenka Kubíčková [2], Kyo-Hoon Ahn [2], Markéta Jarošová [2], Joris More Chevalier [3], Stanislav Cichoň [3], Tomáš Kmječ [4], Jaroslav Kohout [4], Marcus Hans [5], Stanislav Mráz [5], Jochen M. Schneider [5], Esmaeil Adabifiroozjaei [6], Leopoldo Molina-Luna [6], Oliver Gutfleisch [1], Imants Dirba [1**], Karel Knížek [2*]

[1] Functional Materials, Institute of Materials Science, Technical University of Darmstadt, Peter-Grünberg-Str. 16, 64287 Darmstadt, Germany
[2] FZU – Institute of Physics of the CAS, Cukrovarnická 10, 162 00 Praha 6, Czech Republic
[3] FZU – Institute of Physics of the CAS, Na Slovance 1999/2, 182 00 Praha 8, Czech Republic
[4] Faculty of Mathematics and Physics, Charles University, V Holešovičkách 747/2, 180 00 Praha 8, Czech Republic
[5] Materials Chemistry, RWTH Aachen University, Kopernikusstr. 10, 52074 Aachen, Germany
[6] Advanced Electron Microscopy Division, Institute of Materials Science, Technical University of Darmstadt, Peter-Grünberg-Str. 22, Darmstadt 64287, Germany

[*] Corresponding author: knizek@fzu.cz
[**] Corresponding author: imants.dirba@tu-darmstadt.de

## Abstract

A series of thin films Fe$_{4-x}$Ge$_x$N ($x$ = 0 – 1) was fabricated onto MgO substrates by magnetron sputtering with the aim of studying the possible enhancement of the anomalous Nernst effect (ANE), envisaged based on Density Functional Theory (DFT) calculations. The Nernst and Hall effects of the series were systematically analyzed, complemented with resistivity, magnetic, electron microscopy and Mössbauer experiments, and DFT calculations including elastic properties. The Fe$_4$N phase crystallizes in the cubic symmetry with $Pm\bar{3}m$ space group, whereas a small tetragonal distortion is realized in Fe$_{4-x}$Ge$_x$N films for $x$ > 0.35. From the comparison of the experimental isomer shift with DFT calculations, we conclude that Ge occupies the *4b* site in the tetragonal $I4/mcm$ structure. Ferromagnetic T$_C$ decreases rapidly from ~750 K for $x$ = 0 to ~100 K for $x$ = 1. The tetragonal samples with $x$ = 0.8 and 1 display two-component behavior in the Hall and Nernst effects hysteresis loops, which can be analyzed as a sum of positive and negative loops with different saturation fields. This unusual behavior is a product of a combination of several factors. (1) Co-existence of two different crystallographic orientations in the tetragonal thin film, namely with the majority of *c*-axis and minority of *a*-axis normal to the film surface. (2) Opposite sign of the anomalous Hall and Nernst effects for the direction of magnetization along the *a* and c-axis revealed by DFT calculation. (3) The magnetocrystalline anisotropy characterized by an easy *ab*-plane, which is responsible for the different saturation fields for *a* and *c*-axis. The maximum ANE was determined to be 0.9 µV/K for $x$ = 0 at room temperature, and -0.85 µV/K for $x$ = 1 at $T$ = 50 K. The rapid increase of ANE of Fe$_3$GeN from low temperatures indicates that, were it not for its low T$_C$, it could surpass ANE of Fe$_4$N. This observation is consistent with our theoretical assumptions and motivates further research of doped Fe$_4$N for which ANE enhancement is predicted by DFT calculations.

## Introduction

The application relevance of iron nitrides Fe$_x$N ($x$ ≥ 3) results from their favorable magnetic, electrical, and mechanical properties. Since they only consist of iron and nitrogen, both of which are cost-effective, abundant, and non-toxic, they provide cheap, environmentally friendly, and recyclable functional materials [1,2,3]. The first applications of iron nitrides were related to their mechanical and chemical resistance and consisted of creating a coating layer that prevents metal corrosion and increases the mechanical strength of steel. Further applications arose in connection with their excellent magnetic properties with $T_C$ high above room temperature, which expanded their use as e.g. high-density magnetic recording heads, magnetic storage media, strong permanent magnets, and, thanks to their low cytotoxicity, also as magnetic materials for biomedicine [4,5,6,7,8,9]. The phase diagram of iron-nitrogen provides a variety of interesting materials with their magnetic properties tunable over a broad range depending on the nitrogen content. Magnetism correlates with the iron amount, starting from nonmagnetic FeN to ferromagnetic $\gamma'$-Fe$_4$N [6] with high magnetization and even further to $\alpha'$-Fe$_8$N [3,10]. Perhaps the most attention is attracted by the ordered tetragonal superstructure $\alpha''$-Fe$_{16}$N$_2$ [11] due to its unique combination of high saturation magnetization with enhanced magnetocrystalline anisotropy [12], which has been studied for multiple potential applications, such as rare-earth-free permanent magnets [13], two-phase nanocomposite magnets [14], as well as biomedical applications [8]. In addition to magnetic properties, significant transverse thermo-magnetic properties were also discovered in epitaxial Fe$_4$N thin films on various substrates [15,16]. This discovery further expanded the possible application of iron nitrides for thermoelectric energy harvesting or heat-flux sensors [17,18,19]. In our previous work, using ab-initio Density Functional Theory (DFT) combined with Berry curvature calculations, promising systems for anomalous Nernst effect in substituted iron nitrides have been identified [20]. However, a solid solution in the required range is not available for most of the proposed substitutions. For our study, we have chosen the Fe$_{4-x}$Ge$_x$N series as it is one of the few series for which the existence of solid solutions up to $x = 1$ has been proven for bulk samples [21]. This work aimed to prepare the series Fe$_{4-x}$Ge$_x$N up to $x = 1$ in the form of thin films and determine their transverse magneto-transport properties, namely Nernst and Hall effects, and correlate them with theoretical predictions. The study is complemented by a systematic experimental investigation of microstructural, magnetic, and transport properties.

## Methods

Thin films were deposited using magnetron sputtering in a custom-built setup. Fe and Ge targets were co-sputtered in a nitrogen-containing atmosphere onto MgO(001) substrates maintained at 400°C. Ar and N$_2$ were used as sputtering gases, with partial pressures of $4.2\times10^{-3}$ and $0.8\times10^{-3}$ mbar, respectively. The chamber base pressure was around $10^{-7}$ mbar, and the deposition pressure was set to $4.5\times10^{-3}$ mbar. The target-to-substrate distance was fixed at 10 cm for all depositions. To obtain different Fe/Ge compositions, the sputtering power for the Fe target was held constant at 100 W, while the Ge target power was varied between 13 W and 25 W. After the deposition, a protective coating of Al was deposited on selected films using an Al target inside the chamber.

The phase purity and the degree of preferred orientation of the thin films were checked by X-ray diffraction acquired on a powder Bruker D8 Advance diffractometer (CuK$\alpha$ radiation, Lynxeye XE-T position sensitive detector). Lattice parameters were refined using the FullProf package [22]. The surface of the thin films and their stoichiometry were checked by scanning electron microscopy (SEM) and energy-dispersive X-ray spectroscopy (EDX) using Jeol JXA-8230 apparatus. Thin lamellae of the thin films for transmission electron microscopy (TEM) were prepared by focused ion beam (FIB) within a FEI Helios Nanolab 660 dual-beam microscope. The interpretation of the selected-area electron diffraction images was done by the use of the CrysTBox [23]. The chemical composition of the inner region of the films was investigated using atom probe tomography (APT).

The low-temperature (2 – 300 K) measurements of resistivity and Hall effect were performed using the four-probe method of the Electrical Transport Option (ETO) of the Physical Property Measurement System (PPMS, Quantum Design). The magnetic response of the samples within the temperature range 5 – 380 K was measured using a superconducting quantum interference device (SQUID) magnetometer (MPMS-XL, Quantum Design). High-temperature (300 – 850 K) magnetic response was measured using vibrating sample magnetometer (VSM) EZ9 (Microsense) in $N_2$ (<570K) or Ar (>570K) atmosphere. The measurements were carried out in two macroscopically distinct orientations of the external magnetic field, *i.e.* perpendicular and parallel to the sample surface. The Nernst effect was measured using the Thermal Transport Option (TTO) option of PPMS and using a home-made apparatus employing the standard geometry in which the directions of magnetic field $B_z$, thermal gradient $\nabla T_x$, and the resulting electric field $E_y$ are mutually perpendicular. The details about the measurement geometry are discussed in the Results section. The longitudinal and transverse transport coefficients, namely the longitudinal electrical conductivity $\sigma_{xx}$, Seebeck coefficient $S_{xx}$, Hall conductivity $\sigma_{xy}$, and Nernst coefficient $S_{xy}$ are linked with the Nernst conductivity $\alpha_{xy}$ by the formula

$$\alpha_{xy} = \sigma_{xx}S_{xy} + \sigma_{xy}S_{xx} \qquad (1)$$

The ordinary Hall conductivity was calculated using BoltzTraP2 [24]. The details about the calculation of the anomalous Hall conductivity (AHC) $\sigma_{xy}^A$ and the anomalous Nernst conductivity (ANC) $\alpha_{xy}^A$ utilizing the concept of the Berry curvature can be found in [20]. Density-functional theory (DFT) calculations using the Vienna *ab initio* simulation package (VASP) [25] were carried out with a *k* mesh of 20 × 20 × 20 points and a plane-wave cutoff of 600 eV. The projector-augmented wave [26] potentials with the generalized gradient approximation (GGA) [27] were used. For comparison with Mössbauer spectroscopy, we have calculated set of hyperfine parameters for *x* = 1 according to procedures applied in [28].

The mechanical properties were characterized by nanoindentation using a Hysitron TI-900 TriboIndenter. Room-temperature [57]Fe conversion-electron Mössbauer spectra of the $Fe_{4-x}Ge_xN$ samples with *x* = 0.5 and 1 were obtained in constant-acceleration mode using a [57]Co/Rh source; the velocity scale was calibrated against a room-temperature spectrum of an α-Fe foil, and isomer shifts are reported relative to its centroid. More experimental details can be found in Supplemental Material [29].

## Results

### *Phase formation*

The formation of $Fe_{4-x}Ge_xN$ phases (*x* = 0 – 1) in thin films on MgO substrate has been proved by X-ray diffraction (XRD). All the measured thin films exhibit perfect preferential orientation both perpendicular to the film surface and within the film plane. The structure of $Fe_4N$ (*x* = 0) could be inferred from the *fcc* close-packing of γ-Fe by inserting a nitrogen atom into the center of the cube, thus lowering the symmetry to $Pm\bar{3}m$. Another possible description of the structure can be based on the antiperovskite type $ABX_3$, which could be recalled by rewriting the formula as $FeNFe_3$. Within this description, the nitrogen is located in the position of a small cation B in the center of the $BX_6$ ($NFe_6$) octahedron, and the iron is in two crystallographic sites, namely in the position of a big cation A (Fe2) and as ligands X (Fe1) of the $BX_6$ ($NFe_6$) octahedron, see the structure in Fig. 1b. Preferential orientation perpendicular to the film surface is confirmed by standard θ-2θ scan, where only *h*00 (or 00*l*) reflections are detected, see Fig. S1a in Supplemental Material [29]. The preferential orientation within the film plane is confirmed by the azimuthal scan of 113 reflections, see Fig. S1b in Supplemental Material [29], where the zero angle is aligned along the 100 direction of the MgO substrate. It shows 4 separate peaks confirming 4-fold symmetry within the plane. The azimuthal angle of 113 reflection is inclined by 45°, meaning that 100 directions of the thin film and the substrate are parallel.

The symmetry of phases with higher *x* decreased to tetragonal $I4/mcm$ structure having 4× bigger cell with $c_p$:$a_p$ > 1 ($a_p$ and $c_p$ are lattice parameters recalculated to a primitive cell with *Z* = 1), see Fig. 1c. By comparing the experimental isomer shift (IS) of Mössbauer spectral components with our DFT calculations (see Supplemental Material [29]), we can conclude that Ge preferentially occupies the large cation Fe2 site (*4b* in Wyckoff notation). This finding is consistent with results reported for a polycrystalline sample [21]. According to DFT calculations, the electron density at the Ge site is higher than that of a free atom, *i.e.* Ge acquires a negative oxidation state. It is in accordance with Ge preference for the large perovskite site, since Ge anion should have a larger radius than Fe cation. On the other hand, replacing the Fe-Fe metal bond with a stronger Ge-Fe ionic bond reduces the bond length and shrinks the cell volume. The site of Fe1 bonded to N is split to axial Fe1c and in-plane Fe1ab, see the structure in Fig. 1c. From the extrapolation of the $c_p$:$a_p$ ratio, we estimate the border between cubic and tetragonal phase to $x \sim 0.35$, see Fig. 1a. This is a lower $x$ compared to Ref. [21], where the transition to tetragonal structure was observed between $x = 0.75$ and 1. The difference may be connected with two factors: 1) the nitrogen content of $Fe_{4-x}Ge_xN_y$ phases decreases with $x$ less in our samples (see EDX results below) than in [21], where $y \sim 0.55$ for $x = 1$ was determined, 2) mismatch between MgO substrate and $Fe_{4-x}Ge_xN$ induces strain in our films, compared to bulk samples in [21].

The diffraction pattern for tetragonal $x = 0.5$ and 1 shows both 00$l$ and $h$00 peaks (Fig. S1a in Supplemental Material [29]), meaning that both preferential orientations with $a$ and $c$-axis normal to the film surface are present. For $x = 0.5$, the ratio of $c$:$a$ orientation is approx. 40%:60%, for $x = 1$ it is 80%:20%. This observation could be explained by a competition of two factors: 1) tetragonal phases typically grow with $c$-axis normal to the surface because of the growth kinetics and minimization of the surface energy. This tendency is enhanced for a bigger difference between $c$ and $a$-axis. 2) lattice mismatch between $Fe_{4-x}Ge_xN$ phases and MgO substrate, which has a NaCl-type structure with space group $Fm\bar{3}m$ and lattice parameter $a = 4.21$ Å. If we consider that Fe(Ge) is bonded to oxygen in MgO at the interface and compare the distance between oxygens in MgO $d_{O-O} = 2.98$ Å, and the distance between Fe-Ge in $Fe_3GeN$ within $ab$ plane $d_{ab} \sim 2.65$ Å, and within $ac$ plane $d_{ac} \sim 2.70$ Å, there is 11% or 9% mismatch inducing strain on the nitride film. Thus, the orientation with $c$-axis parallel with the film surface could at least partially decrease this strain. For $x = 0.5$, which has smaller $c$:$a$ ratio, the influence of both these factors is comparable. As the ratio $c$:$a$ increases for $x = 1$, the influence of the first factor becomes more dominant and the percentage of $c$-axis normal to the surface increases. Structural data for the $Fe_{4-x}Ge_xN$ films ($x = 0$, 0.5 and 1) are summarized in Table 1.

The rocking-curve scans of 200/002 reflection in Fig. S1c show the degree of preferred orientation of thin films. The preferred orientation is higher for $x = 0$ and 1 (narrower peaks) than for $x = 0.5$. For $x = 0.5$, the large cation site is occupied randomly by Fe and Ge, which introduces a certain degree of disorder into the structure, while for $x = 0$ and 1, the occupancy of each site is close to 100% by individual ions.

Table 1. Lattice parameters and preferred orientation of the $Fe_{4-x}Ge_xN$ thin films.

| $x$(Ge) | space group | $a$ (Å) <br> $a_p$ (Å) [1] | $c$ (Å) <br> $c_p$ (Å) [1] | $V_p$ (Å$^3$) [1] | $c_p : a_p$ [1] | Orientation | |
|---|---|---|---|---|---|---|---|
| | | | | | | 00$l$ | $h$00 |
| 0 | $Pm\bar{3}m$ | 3.804(1) | | 55.03(1) | 1.000 | 100% | |
| 0.5 | $I4/mcm$ | 5.361(1) <br> 3.791(1) [1] | 7.640(2) <br> 3.820(1) [1] | 54.89(1) [1] | 1.008 [1] | 40% | 60% |
| 1 | $I4/mcm$ | 5.311(3) <br> 3.755(2) [1] | 7.762(2) <br> 3.881(1) [1] | 54.73(2) [1] | 1.034 [1] | 80% | 20% |

[1] recalculated to a primitive cell with $Z$ = 1.

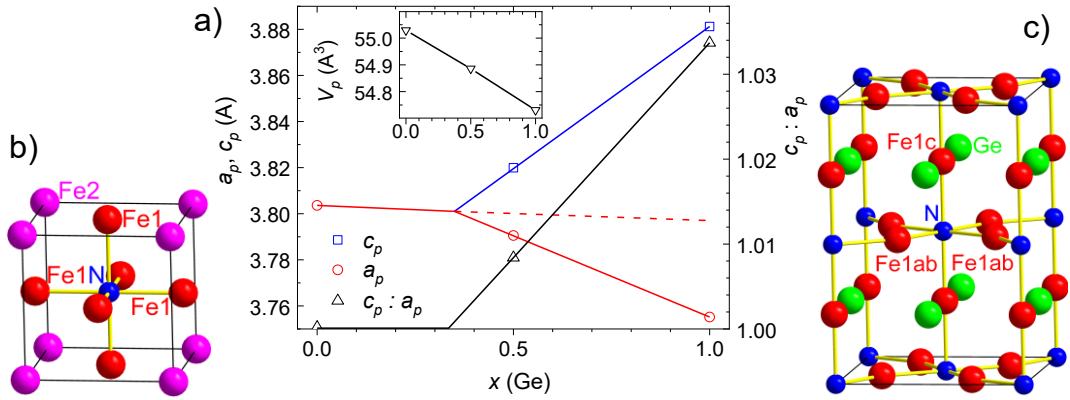

Fig. 1. a) The dependence of the lattice parameters of the primitive cell $a_p$ and $c_p$, ratio of lattice parameters $c_p$:$a_p$, and the volume of the primitive cell $V_p$ on Ge content ($x$). Crystal structure of (b) $Fe_4N$ and (c) $Fe_3GeN$. Displayed atoms are Fe (red/magenta), Ge (green), and N (blue).

## *Microstructure*

The stoichiometry of the films determined by averaging several points in SEM images using EDX is displayed in Table 2. The contents of Fe and Ge are close to the expected composition, while the nitrogen content is slightly lower, and the difference between nominal and determined content increases with $x$. The surface of the film visualized by backscattered electrons using SEM reveals an interesting pattern for $x = 1$, which consists of perpendicular, evenly spaced dark stripes that form an almost regular grid, see Fig. 2. The stripes differ by a lower Ge content of $x \sim 0.85(4)$ from the rest of the sample with $x \sim 0.97(2)$. Let us note that the focus spot of the EDX is larger than the stripe width, so the stripe composition is calculated by comparing the composition of the light area itself and of the mixture of the stripe and the light area. We estimate that the stripes comprise about 15-20% of the thin film volume, which would give overall Ge content $x \sim 0.95$. Since the volume of the stripes corresponds to the fraction of cells with the *a*-axis perpendicular to the film surface, we assign these stripes to this distinct preferential orientation. It would also explain the tendency toward lower Ge content in stripes, since lower $x$ for cells oriented with their *a*-axis perpendicular to the film surface would decrease the difference between lattice parameters *a* and *c* and reduce the possible strain within the film plane, see lattice parameters in Fig. 1. No such stripes were observed for $x = 0$ and $0.5$, since there is no or a small difference between *a* and *c* lattice parameters.

| Table 2. Average composition determined by EDX. | | | |
|---|---|---|---|
| Nominal Ge ($x$) | Fe | Ge | N |
| 0 | 4 | 0 | 0.97(2) |
| 0.5 | 3.51(2) | 0.49(2) | 0.95(3) |
| 1 | 3.05(2) | 0.95(2) | 0.86(3) |

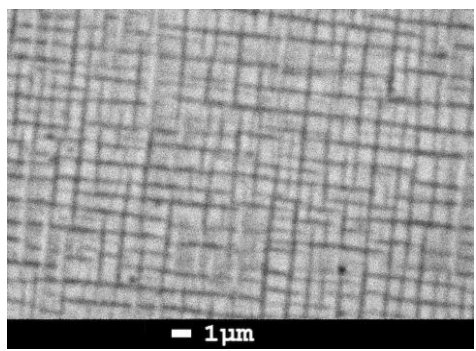

Fig. 2. Backscattered electron image of $x = 1$ film. The Ge content in dark stripes was determined as $x \sim 0.85$, in the light area $x \sim 0.97$.

*Nanostructure*

For deeper insight into the microstructure and composition of the thin films, we employed transmission electron microscopy (TEM) on lamellae cut perpendicular to the film surface using FIB. Fig. 3(a-c) presents a cross-sectional view of the $Fe_4N$ lamella. Overview image (Fig. 3a) shows the stack of layers grown onto MgO substrate with $Fe_4N$ film thickness of about 487(3) nm. Higher magnification image (Fig. 3b) reveals the ~13(2) nm Al capping layer. Perpendicular structural bands within the $Fe_4N$ film can be identified in Fig. 3c. Selected area electron diffraction pattern along [001] zone axis is shown in Fig. 3d. It reveals the cubic symmetry, and the calculated lattice parameter $a = 3.79$ Å agrees well with the value obtained from XRD (3.80 Å). Furthermore, the diffraction pattern taken at the MgO substrate/$Fe_4N$ film interface region shows epitaxial coherent growth, confirming the XRD results, where only $h$00 reflections are detected.

All three samples show band-like formations appearing as regions with varying contrast in the cross-sectional TEM. This effect is particularly illustrative in $Fe_3GeN$ shown in Fig. 4. In Fig. 4a, a low magnification bright-field image of the thin film is pictured. Fig. 4b is the electron diffraction pattern of the thin film, and it shows two different lattice parameters for the tetragonal phase with lattice spacings of approximately 3.8 Å and 3.7 Å corresponding to the crystallographic $c$ and $a$ axis (the second significant digit cannot be extracted from the TEM data because of the experimental error). Using one of the diffraction spots (encircled), a dark-field image is produced and shown in Fig. 4c. It reveals a distinct contrast differences and separation between these regions, suggesting a crystallographic orientation/distortion variation in the $Fe_3GeN$ film. The band-like features are parallel to each other and oriented 45° with respect to the film normal. These structural deviations indicate co-existence of two different crystallographic orientations of the tetragonal phase where parts of the film have $a$-axis orthogonal to the surface in agreement with the XRD results summarized in Table 1. The two regions are separated by anti-phase boundaries, as evident from the high-resolution TEM image shown as Fig. 4d. To investigate the compositional differences, line and point energy-dispersive X-ray spectroscopy (EDX) was performed on these regions. However, no statistically significant variation in Fe and Ge contents could be detected, with the average chemical composition of around 74.4 at.% Fe and 25.6 at.% Ge. This suggests that the observed contrast differences may be related to variations in nitrogen content. Another possible explanation is that the band-like regions change with depth, and since TEM-EDX integrates data over the entire sample thickness, the average composition appears uniform across different areas. To further investigate this phenomenon, atom probe tomography (APT) measurements were conducted and are described in the next section.

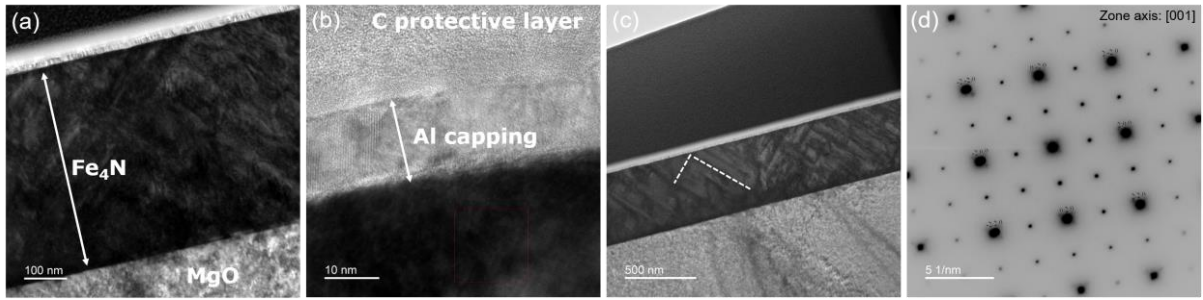

Fig. 3. Transmission electron microscopy results for the Fe$_4$N ($x$ = 0) sample. (a) Overview image showing Fe$_4$N thickness of about 487 nm. (b) Higher magnification image reveals a 13 nm Al capping layer. (c) Perpendicular structural bands are observed in the Fe$_4$N film. (d) Selected area electron diffraction pattern.

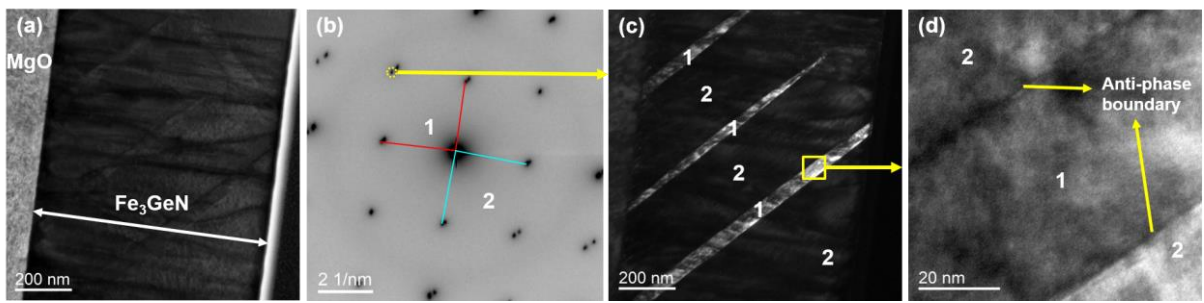

Fig. 4. Transmission electron microscopy results for the Fe$_3$GeN ($x$ = 1) sample. (a) overview low magnification bright-field image. (b) electron diffraction pattern. (c) dark-field image produced from the diffraction spot illustrated in (b). (d) high-resolution TEM image.

## *Composition variations at band-like regions, x = 0, 0.5*

Atom probe tomography (APT) enables spatially-resolved analysis of the chemical composition at the nanometer scale in three dimensions and, therefore, was applied to investigate the micro-/nano-structural features observed in the TEM. It has to be noted for proper interpretation of the results that the Fe–N system exhibits peak overlaps at 28 Da of Fe$^{2+}$ and N$_2^{2+}$. Based on the detected isotopes, the peak at 28 Da has been assigned to Fe$^{2+}$, thus the N content from APT is underestimated. An additional peak overlap of FeN$^{2+}$ and Ge$^{2+}$ is present in the Fe–Ge–N system at 35 Da. This peak has been assigned to Ge$^{2+}$ based on the isotope ratios and, therefore, the Ge content is overestimated. Fig. 5 presents APT results for the Fe$_4$N and Fe$_{3.5}$Ge$_{0.5}$N samples. In the case of Fe$_4$N, the Fe distribution appears uniform throughout the specimen, whereas N shows localized enrichment in band-like regions oblique to the film-substrate interface, which resembles the TEM observations discussed in the previous section. Fig. 5b displays the compositional profiles for Fe and N integrated across the cylindrical volume sketched in Fig. 5a. The cylinder has a diameter of 10 nm and all atoms within the cylinder are counted for 1 nm slices. As explained above, the nitrogen content is consistently underestimated, showing values around 10 at%, although the actual concentration is expected to be closer to 20 at%. Nevertheless, there appears to be a slight reduction in Fe concentration and an increase in N across the band-like features. In the case of the Fe$_{3.5}$Ge$_{0.5}$N sample shown in Fig. 5c, similar 45° inclined bands are observed. However, an interesting point to note is that in contrast to the Fe$_4$N, they are formed by the change in composition of Fe and Ge and not due to a change in nitrogen content, which remains relatively constant throughout the analyzed cylindrical volume, see Fig. 5d.

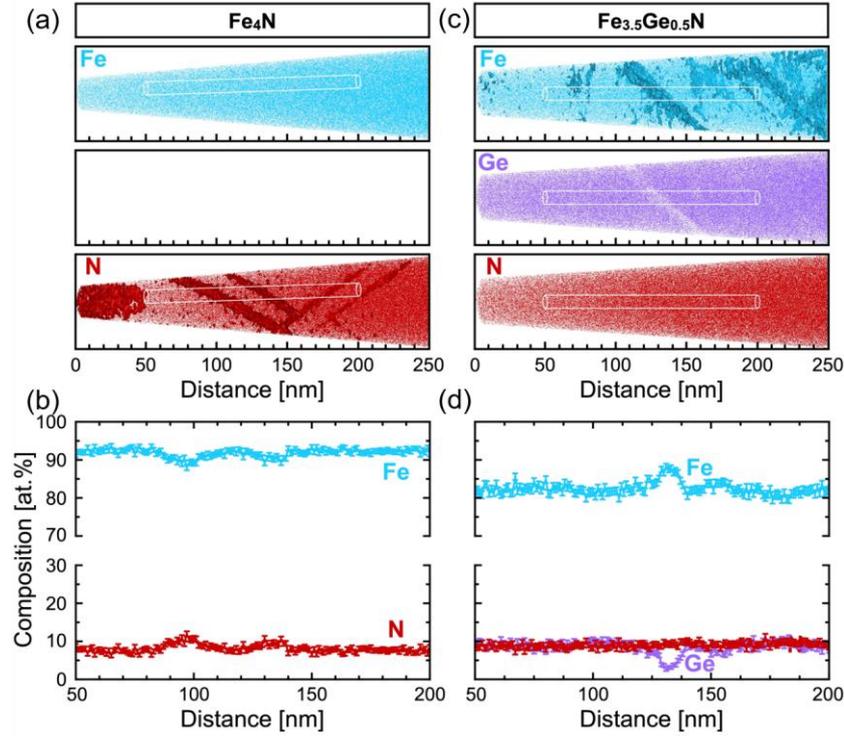

Fig. 5. APT reconstructions for the Fe$_{4-x}$Ge$_x$N samples with (a) $x = 0$ and (c) 0.5. The respective compositional profiles for Fe, Ge and N integrated across a cylindrical volume sketched in (a, c) cylinder with a diameter of 10 nm are shown in (b, d).

In summary of the experimental methods addressing the micro- and nanostructures: Using XRD, both *a*- and *c*-axis crystallographic orientations of the tetragonal $x = 0.5$ and $x = 1$ films were identified, while fraction of *c*-axis orientation increases with *x*. The top-view SEM picture revealed stripe-like features for $x = 1$, and EDX determined different Ge content in and away the stripes, which can be qualitatively based in view of strain energy. The side-view TEM picture revealed similar perpendicular stripes (bands) inclined 45° toward the film surface; they may be ascribed to different crystallographic orientation as well. APT determined a lower Ge content in these stripes consistently with top-view stripes seen by SEM.

### *Magnetic properties*

The magnetic moments calculated by DFT for Fe$_4$N are 2.33 $\mu_B$ for Fe1 bonded to N and 2.93 $\mu_B$ for Fe2 in the large perovskite site, and the total calculated moment is 9.9 $\mu_B$ per formula unit (f.u.). This is higher than the experimental value ~8.5 $\mu_B$ obtained at 10 K (see Fig. 6a). In our DFT calculations of tetragonal Fe$_3$GeN, Fe1 crystallographic site is split to two Fe1c sites along *c*-axis and four Fe1ab sites in the *ab*-plane and Fe2 atom is replaced by Ge in accordance with results of Mössbauer spectroscopy (see Supplemental Material [29]) and Ref. [21]. The total magnetic moment decreases to 2.3 $\mu_B$, not only because the magnetic Fe is replaced by Ge, but also because the magnetic moment of the remaining Fe decreases, namely Fe1c to 1.7 $\mu_B$ and Fe1ab even to 0.3 $\mu_B$, see also density of states (DOS) in Fig. S2 in Supplemental Material [29]. The experimentally determined value of the total moment ~ 2 $\mu_B$ is somewhat lower than the calculated value (see Fig. 7 and Fig. 10a). The critical temperature $T_C$ decreases from 750 K for $x = 0$, to 550 K for $x = 0.5$, and to 100 K for $x = 1$, see Fig. 7. The magnetic moment at 0.5 T decreases similarly from 7.8 $\mu_B$/f.u. for $x = 0$, to 3.3 $\mu_B$/f.u. for $x = 0.5$, and to 1.7 $\mu_B$/f.u. for $x = 1$. For $x = 0$ and $x = 1$, the shape of $M(T)$ loop is similar to conventional magnets, when $M$ decreases quickly close to $T_C$. In contrast, $M$ decreases roughly linearly with increasing temperature for $x = 0.5$, which may be related to the disorder in the structure at the Fe2 site, where Fe and Ge alternate randomly, as opposed to *x* = 0 and 1, where the Fe2 position is fully occupied by Fe or mostly by Ge.

The in-plane hysteresis loops for *x* = 0 are displayed in Fig. 6a,c and for *x* = 0.5 in Fig. 6b,d. For *x* = 0, the curves are almost rectangular with remanent magnetization close to the saturated value, which can be ascribed to perfect crystallinity. For *x* = 0.5, the shape of *M*(*H*) curves is more gradual. The higher coercivity in *x* = 0.5 can be either due to an increase in magnetocrystalline anisotropy caused by the introduction of Ge or random occupation of the site by Fe and Ge. For *x* = 1 the anisotropy decreases, which may nevertheless be caused by a decrease of magnetic interactions overall. Interesting shape of *M*(*H*) curves is for *x* = 1, for which we display both the in-plane and out-of-plane curves in Fig. 10. together with Hall effect loops. In both cases, there is a two-knee feature, while the fields of the knees are shifted for the out-of-plane curves due to demagnetizing effects. The interpretation is discussed in the next section together with transport measurements.

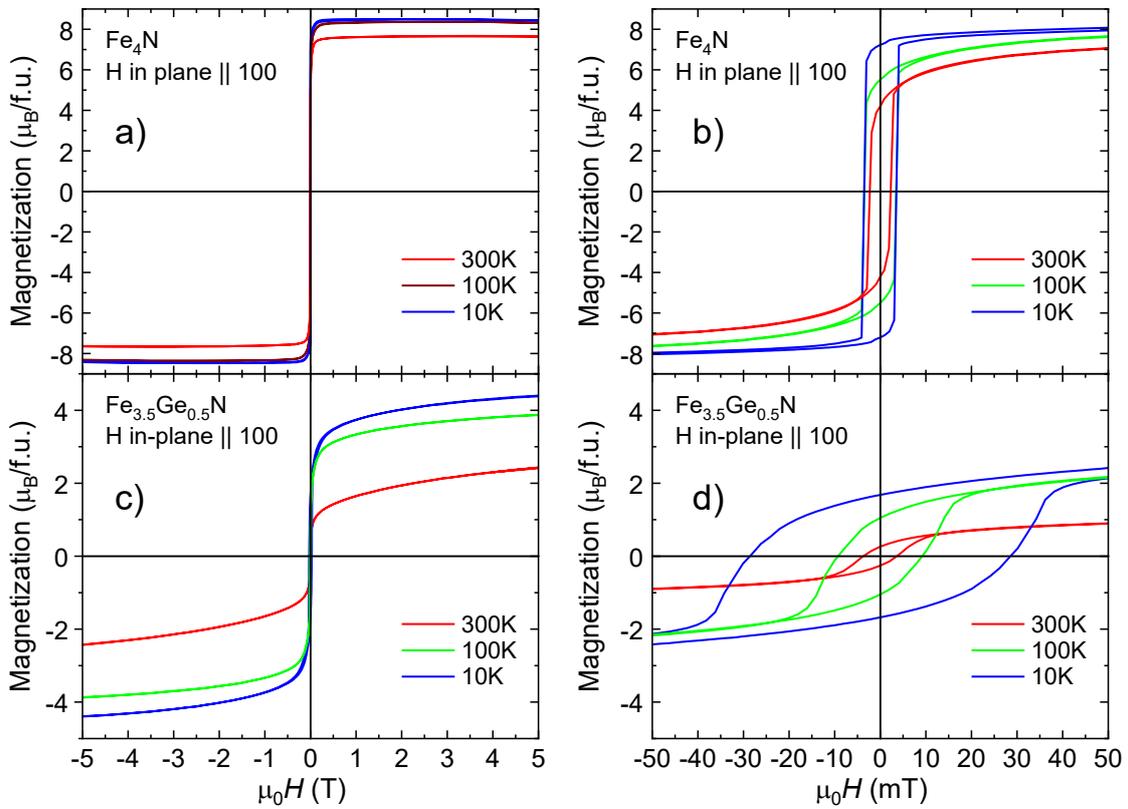

Fig. 6. Hysteresis loops for (a) *x* = 0 and (c) *x* = 0.5, in-plane field. Right panels (b) and (d) display the detailed hysteresis near zero field.

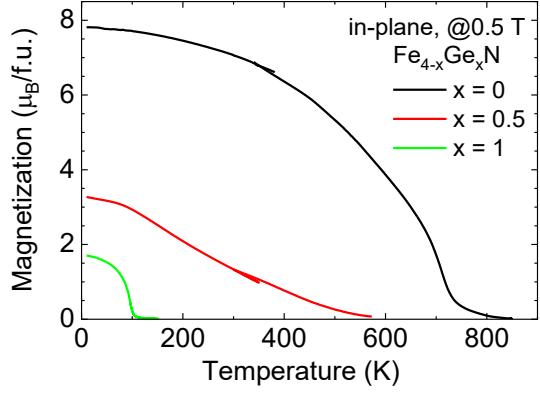

Fig. 7. Temperature dependence of nearly-saturated magnetization for $Fe_{4-x}Ge_xN$ with $x$ = 0, 0.5, and 1. Adding Ge drastically reduces both magnetic moment and Curie temperature, and for non-integer $x$ = 0.5, the shape of the curve is distinct from $x$ = 0 and $x$ = 1 curves.

*Resistivity*

The temperature dependence of resistivity for $x$ = 0, 0.5, and 1 is displayed in Fig. 8. At low temperature, the temperature dependence of resistivity can be fitted by the relation $\rho_0 + AT^n$ with $n = 2.0$ for all $x$. This type of dependence can be related to electron-electron scattering. The residual resistivity $\rho_0$ is 3.1, 82 and 40 µΩ·cm for $x$ = 0, 0.5 and 1, respectively. A high residual-resistivity ratio (RRR) of the $x$ = 0 sample, in our case calculated as $R(300 K)/R(5 K) \sim 27$, indicates a good crystallinity of the thin film. Other literature resistivity data of the $Fe_4N$ thin films show a smaller RRR within the range 3 – 10 [30,31,32,33]. Linear dependence ($n = 1$) is expected at a higher temperature; however, we observe an unusual sublinear ($n = 0.33$) temperature dependence, which cannot be explained by usual terms such as electron-electron, electron-phonon or electron-magnon interactions. Nevertheless, the same sublinear dependence was also observed for $Fe_4N$ thin films by other authors [30,31,32,33]. One explanation might be related to crystallographic defects connected with the lattice mismatch between $Fe_4N$ and MgO. However, the same behavior is observed for thin films deposited on $LaAlO_3$ and $SrTiO_3$, for which the mismatch is much smaller [32]. Similar sublinear dependence was observed in other magnetically ordered metallic phases, *e.g.* $SrIrO_3$ [34], $PdCrO_2$ [35], or $Ni_3Sn_2$ [36] above the temperature of magnetic ordering and attributed to the gradual development of short-range spin correlations, which may reduce the magnetic scattering of the conduction electrons. However, in contrast to these references, in our case we observe the phenomena in the magnetically ordered state. Nevertheless, we propose that in our case, the magnetically ordered state can be perturbed by spin fluctuations that are induced by local magnetic moments associated with N vacancies. For this purpose, we have calculated the electronic structure of a 3×3×3 supercell with one empty N site to simulate N vacancy. This corresponds to $Fe_4N_{0.96}$ stoichiometry, comparable to that determined by EDX. The calculation revealed an enhancement of the magnetic moment for the Fe1 in the vicinity of the N vacancy up to 2.5 µB, compared to Fe1 bonded to N having a magnetic moment 2.33 µB. We propose that the random distribution of N vacancies and the associated random distribution of larger magnetic moments may result in a similar temperature dependence of the resistivity as induced by spin fluctuations.

The RRR ratio $R(300 K)/R(5 K)$ is 4 for $x$ = 1. The lower RRR is caused by two factors, namely the higher residual resistance of 40 µΩ·cm and the change in slope of the dependence at higher temperatures. The temperature dependence of resistivity shows two anomalies; a change of slope around 100 K connected with ferromagnetic transition $T_C$, and a step down around 180 K, which we attributed to a small amount of secondary phase with a lower content of Ge. The sublinear resistivity dependence is observed in the temperature range just below $T_C$ with $n = 0.4$, whereas above $T_C$ the dependence on temperature is roughly linear. For $x$ = 0.5, the RRR ratio is 1.8, which is the smallest value of the series. We relate the low RRR to the disorder in the structure at the Fe2 site (the large position of the anti-perovskite structure), where Fe and Ge alternate randomly, as opposed to *x* = 0 and 1, where the Fe2 position is fully occupied by Fe or mostly by Ge. The sublinear behavior is not evident for *x* = 0.5, but it is probably masked by enhanced resistivity due to the site disorder.

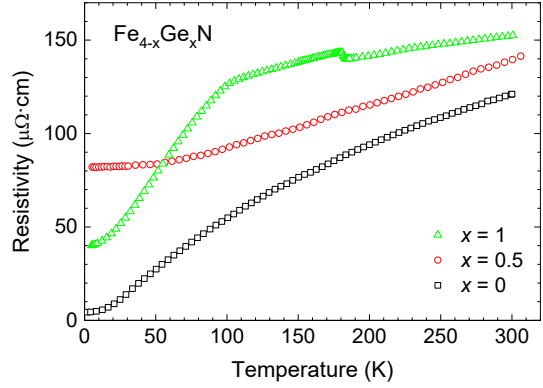

Fig. 8. Temperature dependence of resistivity for $Fe_{4-x}Ge_xN$ thin films with *x* = 0, 0.5 and 1.

### *Hall effect, x = 0, 0.5*

The temperature dependences of ordinary and anomalous Hall and Nernst effects extracted from the experimental hysteresis loops at individual temperatures are displayed in Fig. 9 for *x* = 0, 0.5, 0.8 and 1. The measurement geometry for the Hall effect is displayed in Fig. 11a and for Nernst effect in in Fig. 11c. Nernst effect measurement using PPMS (Fig. 11b) was only made for a few selected temperatures to determine the temperature gradient.

Ordinary Hall effect (OHE) for *x* = 0 is positive at room temperature, the dependence on temperature is linear and it smoothly changes sign from positive to negative crossing zero at around 75 K and then it changes back to a positive value around 20 K. This behavior reflects the properties of a multi-carrier system, possessing a complicated band structure, where we can identify five bands crossing the Fermi level, see Fig 3. in Ref. [20]. The bands have both hole and electron characters (negative and positive effective mass, respectively). The positive sign of the OHE indicates that the holes dominate the transport properties. The Boltztrap calculation determined a positive Hall sign in accordance with the experiment but failed to reproduce the sign change at low temperatures. Boltztrap uses the calculated band structure, so its output captures the number of carriers and their effective mass for all bands. However, its critical approximation lies in a single scattering time for all carrier types. We estimate that the sign change is related to the different temperature dependence of the scattering times of individual carriers. OHE for *x* = 0.5 is also positive at room temperature and linearly decreases towards zero at low temperatures. A higher value of the OHE compared to *x* = 0 indicates lower apparent carrier concentration. Anomalous Hall effect (AHE) for *x* = 0 is also positive at room temperature, decreases almost linearly with decreasing temperature, and tends towards zero at the lowest temperature. In contrast to OHE, no change of sign is observed. A relation linking anomalous Hall resistivity $\rho_{xy}^{AH}$ and longitudinal resistivity $\rho_{xx}$ has been devised in the form $\rho_{xy}^{AH} \sim \lambda \rho_{xx}^n$ with $n$ and $\lambda$ as nondimensional exponent and prefactor [37]. The exponent $n$ in this power-law relation is related to different mechanisms of the AHE. The intrinsic models that only depend on the ideal band structure and are independent of scattering include interband effect and the Berry curvature mechanisms. The models based on extrinsic mechanisms include side jump and skew scattering. The intrinsic and side jump mechanisms can be modeled with the exponent $n = 2$, while the skew-scattering mechanism predicts $n = 1$. The relation $\rho_{xy}^{AH} \sim \lambda \rho_{xx}^n$ cannot be well fitted over the whole range for *x* = 0, but the fit is significantly improved if it is applied for two separate ranges. For *T* < 70 K the fit gives $n = 1.96$, whereas for *T* > 70 K the best fit is obtained for $n = 1.01$, see Fig. S3 in Supplemental Material [29]. This would suggest that the AHE mechanism changes with temperature. The temperature dependence of AHE for $x$ = 0.5 is rather constant, slightly increasing with lowering temperature. The above-discussed relation is not applicable in this case.

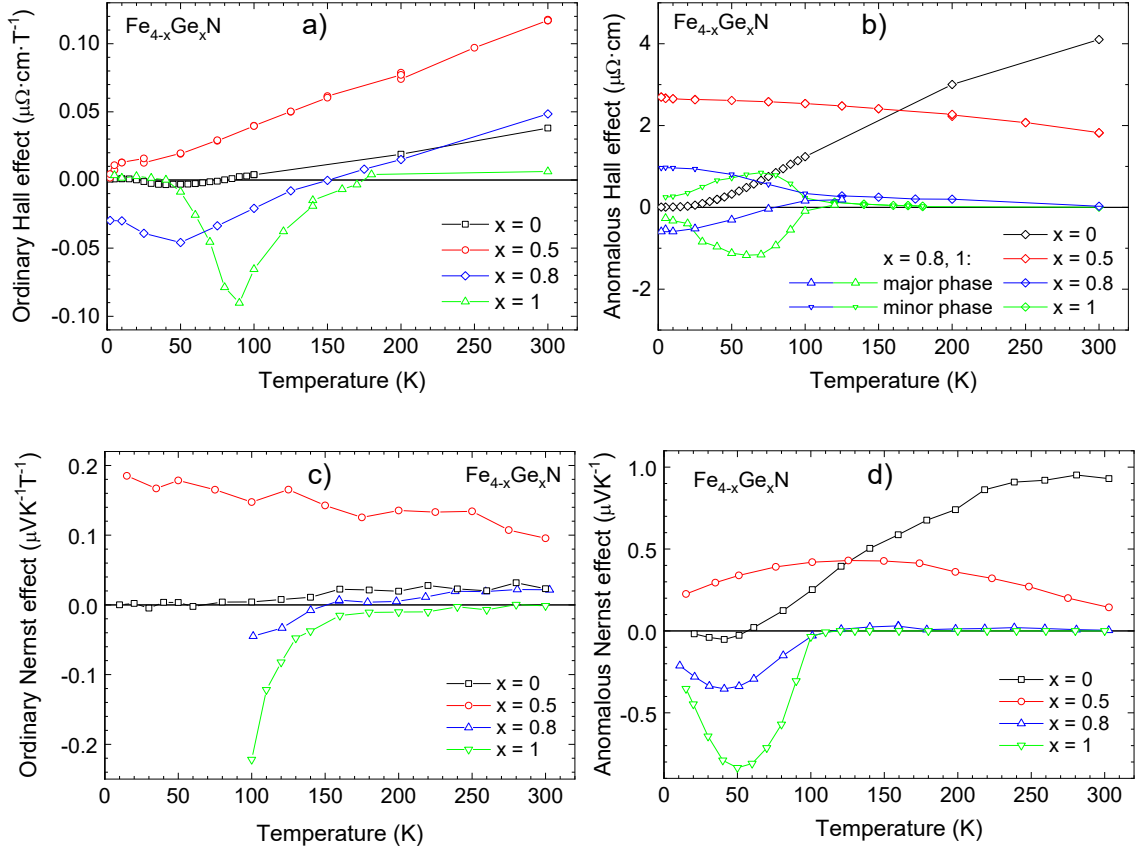

Fig. 9. Temperature dependence of (a) ordinary Hall, (b) anomalous Hall, (c) ordinary Nernst, and (d) anomalous Nernst effects for $x$ = 0, 0.5, 0.8 and 1.

## *Nernst effect, x = 0, 0.5*

The Nernst effect was measured using the home-made apparatus, where the magnetic field and the measured electric field are in the plane of the thin film and the temperature gradient is perpendicular to the thin film, see Fig. 11c. The Nernst effect loops for $x$ = 0, 0.5, and 1 are shown in Fig. S4 in Supplemental Material [29]. The hysteresis loops are comparable with in-plane magnetic measurements. In this configuration, the temperature gradient cannot be measured separately for the thin film and the substrate, but the sum of both contributions is determined. The total gradient is divided proportionally according to the thermal resistances of the film and substrate. However, the thermal resistance of the materials involved may not be known or may be different than the literature values for the specific samples used. Therefore, the determination of the thermal gradient of the film itself is uncertain. A more accurate way is to make additional measurements using the configuration shown in Fig. 11b, which is available in PPMS apparatus. In this case, the temperature gradient is applied in the plane of the film and can be accurately determined. Since the temperature evolution of the thermal resistances of the film and substrate is generally different, several selected temperatures were measured. The hysteresis loops are comparable with out-of-plane magnetic measurements. In agreement with Berry curvature calculation (see Fig. 12), the ANE for $x$ = 0 is positive. The maximum at room temperature is 0.9 µV/K. It decreases by lowering the temperature and changes the sign around 55 K, and after a minimum around 40 K tends to zero at the lowest temperature. In the case of $x$ = 0.5,

ANE is positive with a maximum around 100 K and tends to have a positive value at the lowest temperature.

### *Hall and Nernst effects, x = 1*

Magnetization loops for *x* = 1 exhibit a two-component character, see Fig. 10. The in-plane loop (Fig. 10a) grows quickly to an almost saturated value in the low-field region, followed by a slow linear increase, saturating at ~2 T. The out-of-plane curve (Fig. 10b) exhibits a similar behavior, but the contribution of the slowly-saturating part to the total moment is much larger. The two-component feature is more prominent in the Hall effect measurements (Fig. 10c), corresponding to the out-of-plane setup (Fig. 11a), because the two components have opposite AHE signs. The example of decomposition of the loop into the contribution of OHE and two contributions of AHE for *T* = 40 K is displayed in Fig. 11a. A similar shape with two ANE contributions of opposite signs is also observed for Nernst effect hysteresis loops, see analysis in Fig. 11b.

Similar two-component character of the Hall effect was observed in Ref. [39], where the authors reported the observation of AHE in $(Bi,Mn)_2Se_3$ thin films and showed that the sign of AHE changes from positive to negative as the Mn concentration is increased. The positive and negative AHE were found to coexist in a crossover regime and could be distinguished due to significantly different saturation fields. The authors assigned positive AHE with higher saturation field to the bulk states and negative AHE with lower saturation field to the surface states.

In our case, we propose that the two-component shape of the Hall and Nernst loops is caused by different signs of Hall and Nernst effects of the majority and minority phases forming the dark stripes observed by SEM, see Fig. 2. First, we tested the possibility that the behavior is related to the lower Ge content of the stripes. For this purpose, we have prepared thin films with *x* = 0.8 and measured Hall and Nernst effects. In the case of this scenario, we would see opposite signs of the Hall effect for *x* = 0.8 and for *x* = 1. However, we also observed similar two-component behavior for *x* = 0.8, but with different temperature dependence, see Fig. 9b. We then focused on the possibility that the interpretation is related to distinct AHE and ANE signs along different crystallographic directions of $Fe_3GeN$ tetragonal cell. For this purpose, we have calculated anomalous Hall and Nernst conductivities using DFT and the Berry curvature mechanism. The calculations indeed confirmed opposite signs of the effects calculated for magnetization along [100] and [001] directions, both for AHC and ANC, see Fig. 12.

Therefore, we propose that the two-component shape of the hysteresis loops is related to the two orientations of tetragonal cells in thin films (see XRD and SEM sections) and opposite signs of the anomalous Hall and Nernst effect for magnetization along *c* and *a*-direction. According to XRD, the majority (~80%) of the *x* = 1 thin film is oriented with *c*-axis perpendicular to the surface. When measuring the Hall effect, the electric current and the measured voltage are in the plane, see Fig. 11a. This means that AHE is observed for the magnetization direction ||[001], which, according to Berry curvature calculations, should be negative, see Fig. 12a. The minor part of the thin film is oriented with *c*-axis parallel to the surface, so for this part the AHE with magnetization ||[100] or [010] is detected, which is positive, see Fig. 12a. This explanation is also supported by the measurement of the Nernst effect for *x* = 1 using PPMS, which is measured in the same geometry, just the electric current is replaced by a temperature gradient. According to the Berry curvature calculations, the ANE measured with magnetization along *c*-direction is positive and ANE measured with magnetization along *a*-direction is negative, *i.e.* opposite signs compared to AHE, see Fig. 12b. This is consistent with the experiment, since we also observe a two-component character of the hysteresis loop for the Nernst effect, but with opposite signs, see Fig. 11b.

Measurements of the Nernst effect performed in the home-made apparatus differ in the experimental setup, namely in the interchange of directions of $\nabla T$ and $M$, and also in a lower attainable field, see Fig. 11c. In this setup, the ANE ratio for *M*||*c* and *M*||*a* is substantially different. While for the

PPMS setup it should correspond to the ratio of crystallographic orientations of approximately 80:20, in the home-made apparatus, this ratio should be about 10:90 (assuming that for the orientation with *a*-axis perpendicular to the surface, the portion of the *c*-axis orientation is the same for both directions). The negative value of ANE for *M*||*a* should therefore be approximately 4.5× greater, and the positive value for *M*||*c* approximately 8× smaller. Since the measured magnetic field range is smaller than the saturation field for *M*||*c*, the two-component behavior is not visible in this measurement, and the contribution of ANE with *M*||*c* only exhibits as a slight linear decrease in the Nernst signal, which is essentially indistinguishable from ONE contribution. Therefore, ONE contribution was not determined for *x* = 1 below $T_C$ (Fig. 9c). As regards the absolute value of the negative ANE contribution, it is about 3× greater than in the PPMS setup, which is less than according to the previous estimate. This discrepancy may be explained *e.g.* by uncertainty in estimating the ratio of different crystallographic orientations, or deviations in determining the temperature gradient in individual measurements, but overall, we can consider it as a good agreement.

In order to explain different saturation fields for *a* and *c*-direction, we have calculated the energy differences among various magnetization directions by DFT calculations including spin-orbit coupling. It revealed that the *M*||[100] configuration has lower energy by 0.272 meV/f.u. than *M*||[001]. This energy difference corresponds to an anisotropy field of 2.3 T, assuming a saturated magnetic moment of 2 $\mu_B$ for 40 K, see Fig. 10. The observed saturation field of the unfavorably oriented component around 2 T is in good agreement with the calculated energy difference. Moreover, also magnitudes of magnetization can be explained within this scenario: For the in-plane measurement (Fig. 10a), the majority orientation with large contribution is saturated at low fields, while for the out-of-plane case (Fig. 10b), the minority orientation is saturated at low fields, and majority of the signal is gained between 0.3 and 2.2 T (between two dashed lines in Fig. 10b).

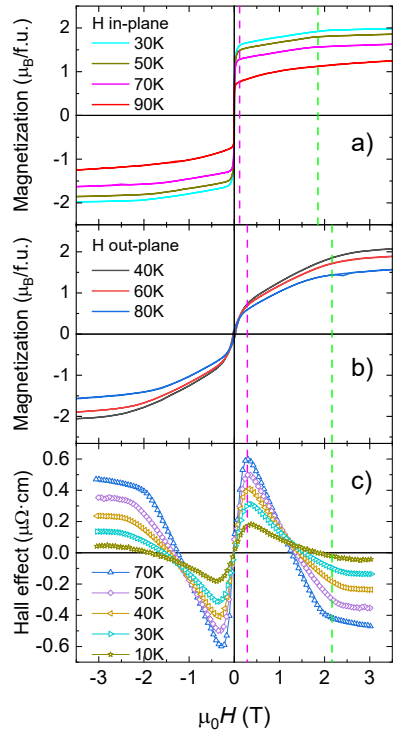

Fig. 10. (a) In-plane and (b) out-of-plane magnetization curves for $Fe_3GeN$ as compared to (c) Hall effect loops. Both in-plane and out-of-plane magnetization curves show 2-step behavior. Critical fields are higher for the out-of-plane case due to demagnetization field. Hall effect loops correspond to the out-of-plane configuration, and two-step feature is even more prominent, because the two contributions have opposite signs of AHE, corresponding to different critical fields.

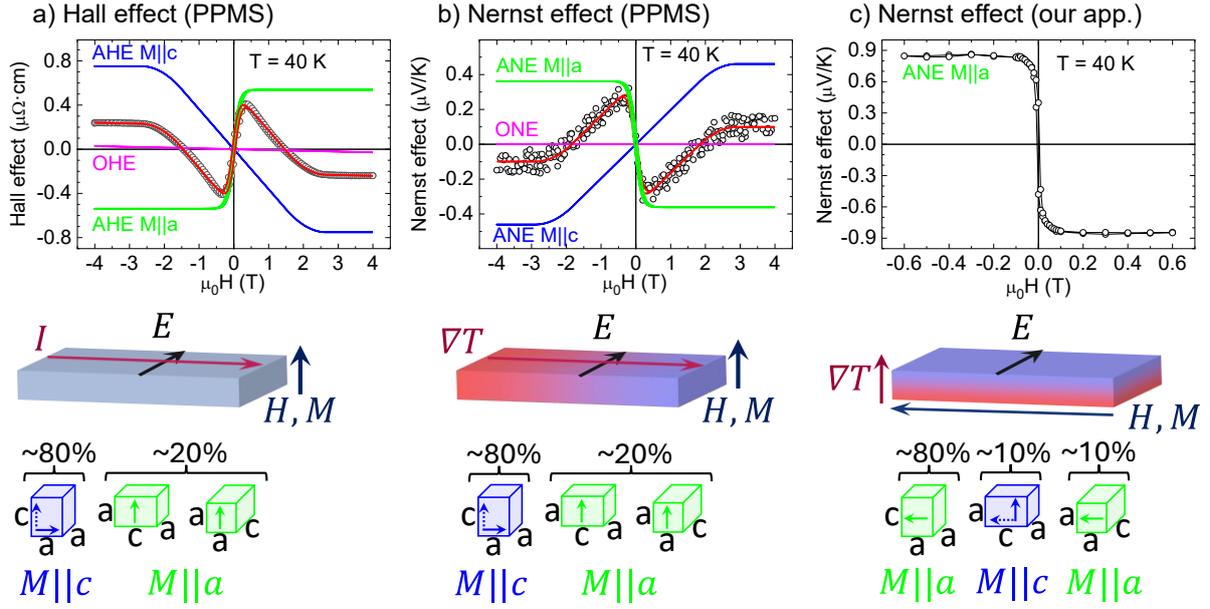

Fig. 11. Hall and Nernst effect loops for tetragonal Fe$_3$GeN at $T$ = 40 K with simulation of two-component AHE/ANE and linear OHE/ONE. Data and measurement geometry for (a) Hall effect in PPMS apparatus, (b) Nernst effect in PPMS, and (c) Nernst effect in home-made apparatus. Black circles: experimental data (corrected for constant and symmetric contributions). Magenta lines: OHE or ONE component. Blue curves: AHE or ANE component for magnetization parallel to *c*-axis. Green curves: AHE or ANE component for magnetization parallel to *a*-axis (see corresponding calculations in Fig. 12). Red curves: sum of all contributions, fitting the experimental data. Displayed tetragonal unit cells show the distribution of crystallographic orientation in the film as determined by XRD. Full arrows in the unit cells indicate magnetization in the direction of the easy *a*-axis, dotted arrows denote magnetization rotated to the direction of hard *c*-axis.

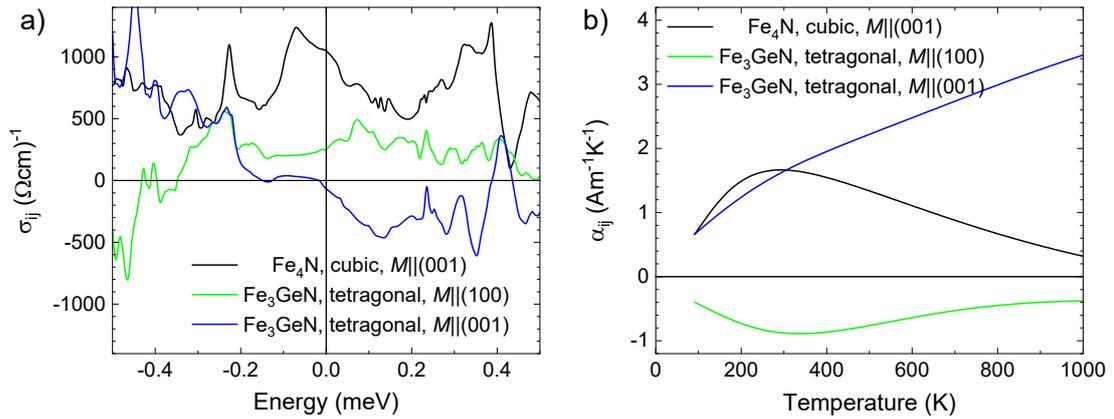

Fig. 12. Comparison of the calculated intrinsic (a) AHC ($\sigma_{ij}$) and (b) ANC ($\alpha_{ij}$), for cubic Fe$_4$N and tetragonal Fe$_3$GeN for magnetization along *c*-axis $M$||[001] and along *a*-axis $M$||[100]. Calculation was made using Berry curvature mechanism.

## Conclusions

Series of $Fe_{4-x}Ge_xN$ ($x$ = 0 – 1) thin films was grown onto MgO substrates by magnetron sputtering. All the measured thin films exhibit epitaxial growth. The $Fe_4N$ ($x$ = 0) phase crystallizes in the cubic $Pm\bar{3}m$ space group, whereas a small crystallographic tetragonal distortion toward $I4/mcm$ symmetry is realized in $Fe_{4-x}Ge_xN$ films, with the estimated border between cubic and tetragonal phase about $x \sim 0.35$. From the comparison of Mössbauer spectroscopy and DFT calculation results, we conclude that Ge occupies the *4b* site in the tetragonal structure. All samples are ferromagnetic with Curie temperature decreasing rapidly from 750 K for $x$ = 0 to 100 K for $x$ = 1.

An unusual sublinear temperature dependence of resistivity was observed for $x$ = 0 and 1, which cannot be explained by standard electron interactions. A similar type of dependence was identified for some magnetically ordered metallic phases, where it was attributed to the gradual development of short-range spin correlations above $T_C$. We propose that in our case, where the sublinear dependence is observed below $T_C$, the random distribution of N vacancies and the associated random distribution of larger magnetic moments perturb the magnetically ordered state, which may result in a similar temperature dependence of the resistivity as that induced by spin fluctuations. The unusual behavior was not clear for the $x$ = 0.5 sample differing in several aspects from $x$ = 0 and $x$ = 1, which was ascribed to compositional disorder. Besides resistivity, for which $x$ = 0.5 sample showed lowest RRR, the anomalies were observed in magnetization and mechanical measurements (see Supplemental Material [29]) compared to $x$ = 0 and $x$ = 1 samples.

The anomalous Hall and Nernst effects are positive for $x$ = 0 and 0.5. The maximum ANE signal was determined 0.9 μV/K for $x$ = 0 at room temperature. The relation $\rho_{xy}^{AH} \sim \lambda \rho_{xx}^n$, in which the exponent $n$ corresponds to the type of AHE mechanism, cannot be reasonably fitted over the entire measured temperature range of 2 – 300 K. However, the fit can be applied to two separate ranges, namely for $T$ < 70 K the resulting exponent is $n = 1.96$, while for $T$ > 70 K the fit gives $n = 1.01$. This would suggest that the AHE mechanism changes with temperature for $x$ = 0. The maximum ANE signal for $x$ = 1 is -0.85 μV/K at $T$ = 50 K. The rapid increase of ANE of $Fe_3GeN$ from low temperatures indicates that it would surpass ANE of $Fe_4N$ if it were not limited by the low $T_C$. This observation is consistent with our theoretical assumptions and encourages further attempts of syntheses of doped $Fe_4N$ where ANE enhancement is expected according to DFT combined with Berry curvature calculations.

The hysteresis loops of the Hall and Nernst effects exhibit unusual two-component shapes, which can be described as the sum of positive and negative loops with different saturation fields. We explained this observation by the presence of two different crystallographic orientations in thin films, with the majority/minority of the sample oriented with the *c/a*-axis perpendicular to the film surface, and opposite signs of the respective phenomena for different directions. The opposite signs of the effects were confirmed by DFT calculations in combination with the Berry curvature method. The presence of two different orientations was revealed by XRD and is consistent with the observation of stripes forming an almost regular orthogonal structure with varying Ge content by SEM, the detection of bands with different crystallographic orientation by TEM, as well as variations of Ge content in these bands determined by APT. The formation of these stripes is driven by the tendency to partially reduce the lattice mismatch between the film and the substrate.

## Acknowledgments

Authors thank O. Heczko for help with high-temperature magnetization measurements. This work was supported by Project ID 22-10035K of the Czech Science Foundation, and by Project ID 471878653 of the Deutsche Forschungsgemeinschaft (DFG, German Research Foundation). We acknowledge the Operational Program Research, Development and Education financed by the European Structural and Investment Funds and by the Ministry of Education, Youth and Sports (MEYS) of the Czech Republic, Grant No. CZ.02.01.01/00/22 008/0004594 (TERAFIT). Computational resources were provided by the MEYS of the Czech Republic through the e-INFRA CZ, ID:90254. The experiments were performed at Materials Growth & Measurement Laboratory (MGML) supported within the program of Czech Research Infrastructures, project no. LM2023065.## References

1. S. Isogami and Y.K. Takahashi, Antiperovskite Magnetic Materials with 2p Light Elements for Future Practical Applications, Adv. Electron. Mater. 9, 2200515 (2023). doi:10.1002/aelm.202200515
2. T. K. Kim and M. Takahashi, New magnetic material having ultrahigh magnetic moment, Appl. Phys. Lett. 20, 492 (1972). doi:0.1063/1.1654030
3. I. Dirba, P. Komissinskiy, O. Gutfleisch, and L. Alff, Increased magnetic moment induced by lattice expansion from $\alpha$-Fe to $\alpha'$-Fe$_8$N, J. Appl. Phys. 117, 173911 (2015). doi:10.1063/1.4919601
4. S. Bhattacharyya, Iron nitride family at reduced dimensions: A review of their synthesis protocols and structural and magnetic properties, J. Phys. Chem. C 119, 1601 (2015). doi:10.1021/jp510606z
5. J.M.D. Coey, and P.A.I. Smith, Magnetic nitrides, J. Magn. Magn. Mater. 200, 405 (1999). doi: 10.1016/S0304-8853(99)00429-1
6. I. Dirba, M. Baghaie Yazdi, A. Radetinac, P. Komissinskiy, S. Flege, O. Gutfleisch, and L. Alff, Growth, structure, and magnetic properties of γ'-Fe$_4$N thin films, J. Magn. Magn. Mater. 379, 151 (2015). doi:10.1016/j.jmmm.2014.12.033
7. Y. Chen, D. Gölden, I. Dirba, M. Huang, O. Gutfleisch, P. Nagel, S. Schuppler M. Merz, G. Schütz, L. Alff, and E. Goering, Element-resolved study on the evolution of magnetic response in Fe$_x$N compounds, J. Magn. Magn. Mater. 498, 166219 (2020). doi:10.1016/j.jmmm.2019.166219
8. I. Dirba, C. K. Chandra, Y. Ablets, J. Kohout, T. Kmječ, O. Kaman, and O. Gutfleisch, Evaluation of Fe-nitrides, -borides and -carbides for enhanced magnetic fluid hyperthermia with experimental study of $\alpha''$-Fe$_{16}$N$_2$ and $\epsilon$-Fe$_3$N nanoparticles, J. Phys. D: Appl. Phys. 56, 025001 (2023). doi:10.1088/1361-6463/aca0a9
9. Y. Ablets, L. Kubickova, A. Chanda, I. Orue, D. Koch, S. Kim, B. Zhao, S. Najma, S. Forg, E. Adabifiroozjaei, L. Molina-Luna, H. Kang, J.P. Hofmann, H. Zhang, T. Kmjec, J.Angel Garcia, F. Plazaola, R.von Klitzing, W. Donner, H. Srikanth, O. Gutfleisch, I. Dirba, Gram-Scale Synthesis of Fe$_3$N Nanoparticles via a One-Pot Thermal Decomposition Route: Implications for Magnetic Fluid Hyperthermia Applications, ACS Appl. Nano Mater, *available online* (2025). doi:10.1021/acsanm.5c03148
10. G. d'Andrea, S. Zhou, U. Kentsch, M. Major, P. Rani, G. Gkouzia, B. Zhao, Y. Ablets, I. Dirba, H. Zhang, L. Alff, Improved lattice elongation for Fe$_8$Nx (x > 1) thin films prepared via nitrogen ion implantation, AIP Advances 15, 035244 (2025). doi:10.1063/9.0000902
11. I. Dirba, C.A. Schwobel, L.V.B. Diop, M. Duerrschnabel, L. Molina-Luna, K. Hofmann, P. Komissinskiy, H.-J. Kleebe, O. Gutfleisch, Synthesis, morphology, thermal stability and magnetic properties of $\alpha''$-Fe$_{16}$N$_2$ nanoparticles obtained by hydrogen reduction of γ-Fe$_2$O$_3$ and subsequent nitrogenation, Acta Mater. 123, 214 (2017). doi:10.1016/j.actamat.2016.10.061
12. H. Zhang, I. Dirba, T. Helbig, L. Alff, and O. Gutfleisch, Engineering perpendicular magnetic anisotropy in Fe via interstitial nitrogenation: N choose K, APL Mater. 4, 116104 (2016). doi:10.1063/1.4967285
13. J. Wang, Environment-friendly bulk Fe$_{16}$N$_2$ permanent magnet: Review and prospective, J. Magn. Magn. Mater. 497, 165962 (2020). doi:10.1016/j.jmmm.2019.165962
14. I. Dirba, M. Mohammadi, F. Rhein, Q. Gong, M. Yi, B. Xu, M. Krispin, and O. Gutfleisch, Synthesis and magnetic properties of bulk $\alpha''$-Fe$_{16}$N$_2$/SrAl$_2$Fe$_{10}$O$_{19}$ composite magnets, J. Magn. Magn. Mater. 518, 167414 (2021). doi:10.1016/j.jmmm.2020.167414
15. S. Isogami, K. Takanashi, and M. Mizuguchi, Dependence of anomalous Nernst effect on crystal orientation in highly ordered γ'-Fe$_4$N films with anti-perovskite structure, Appl. Phys. Express 10, 073005 (2017). doi:10.7567/APEX.10.073005
16. K. Ito, J. Wang, Y. Shimada, H. Sharma, M. Mizuguchi, and K. Takanashi, Enhancement of the anomalous Nernst effect in epitaxial Fe$_4$N films grown on SrTiO$_3$(001) substrates with oxygen deficient layers, J. Appl. Phys. 132, 133904 (2022). doi:10.1063/5.0102928
17. Y. Sakuraba, Potential of thermoelectric power generation using anomalous Nernst effect in magnetic materials, Scr. Mater. 111, 29 (2016). doi:10.1016/j.scriptamat.2015.04.034
18. T. Higo, Y. Li, K. Kondou, D. Qu, M. Ikhlas, R. Uesugi, D. Nishio-Hamane, C. L. Chien, Y. Otani, and S. Nakatsuji, Omnidirectional control of large electrical output in a topological antiferromagnet, Adv. Funct. Mater. 31, 2008971 (2021). doi:10.1002/adfm.202008971
19. A. Sakai, S. Minami, T. Koretsune, T. Chen, T. Higo, Y. Wang, T. Nomoto, M. Hirayama, S. Miwa, D. Nishio-Hamane, F. Ishii, R. Arita, and S. Nakatsuji, Iron-based binary ferromagnets for transverse thermoelectric conversion, Nature (London) 581, 53 (2020). doi:10.1038/s41586-020-2230-z

# Supplementary information

## Experimental

Thin lamellae of $Fe_4N$, $Fe_{3.5}Ge_{0.5}N$ and $Fe_3GeN$ films for transmission electron microscopy (TEM) were prepared by focused ion beam (FIB) within a FEI Helios Nanolab 660 dual-beam microscope. The region of interest was covered with an Al protective layer of 8 µm length and 2 µm width. First ~50 nm of C were deposited with the electron beam, followed by ~1.5 µm of Pt using Ga ions. The lift-outs were attached to Omniprobe holders and thinning was done by milling a 4 µm wide window with a final lamella thickness of ~50 nm. High-resolution transmission electron microscopy was performed using a JEOL JEM 2100F with an acceleration voltage of 200 kV. ImageJ software was used for image processing and analysis. The interpretation of the selected-area electron diffraction images was done by the use of the CrysTBox [40].

The chemical composition of the inner region of the films was investigated using atom probe tomography (APT). This technique combines mass spectrometry as well as projection microscopy and is based on field evaporation [41]. A voltage is applied between a needle-shaped specimen and an electrode, while control of the field evaporation process is obtained by additional application of voltage or laser pulses. Atoms are eventually evaporated and ionized within the electric field. These ions are accelerated towards a position-sensitive detector, and a mass spectrum is obtained through a time-of-flight measurement. Based on the hit position at the detector, the original position of an atom within the specimen before field evaporation can be reconstructed using projection algorithms. This makes APT a powerful tool enabling identification of changes of the chemical composition at materials defects such as interfaces [42] or grain boundaries [43].

Needle-shaped specimens were fabricated by FIB techniques in the FEI Helios Nanolab 660 dual-beam microscope. Ga ions were employed at 30 kV and a standard protocol has been followed [44]. The specimen preparation was finished with a low voltage cleaning at 5 kV and 40 pA for 30 seconds. APT specimens were transferred to a CAMECA local electrode atom probe (LEAP) 4000X HR and the exposure time to atmosphere during transfer was < 3 min. Field evaporation was assisted by thermal pulsing, employing an ultraviolet laser. A relatively low pulse energy of 10 pJ was chosen in order to maximize the electric field strength and thereby enhance the measurement accuracy [45]. The laser pulse frequency was 200 kHz, while the base temperature of 60 K and detection rate of 0.5% were set. Ranging of the mass spectra and reconstruction of the atomic positions has been carried out in the AP Suite 6.2 software package.

The mechanical properties were characterized by nanoindentation using a Hysitron TI-900 TriboIndenter. Quasistatic measurements were conducted with a 100 nm Berkovich diamond tip and the load-displacement curves were evaluated according to the method from Oliver and Pharr [46]. Load-controlled measurements were carried out with maximum load of 1500 µN, resulting in contact depths of 53 to 58 nm for $Fe_4N$, 45 to 54 nm for $Fe_{3.5}Ge_{0.5}N$ and 47 to 56 nm for $Fe_3GeN$. The tip area function was determined from a fused silica reference sample.

Room-temperature $^{57}Fe$ conversion-electron Mössbauer spectra of the $Fe_{4-x}Ge_xN$ samples with $x$ = 0.5 and 1.0 were obtained in a constant-acceleration mode using a $^{57}Co/Rh$ source. The sample plane was perpendicular to the direction of the $\gamma$-rays. The velocity scale was calibrated against a room-temperature spectrum of an $\alpha$-Fe foil, and isomer shifts are reported relative to its centroid. Spectral analysis was performed using the Confit software [47] ($x$ = 1) or the relaxation model based on the work of Blume and Tjon [48] implemented in the Recoil [49] program ($x$ = 0.5).

## Results

*X-ray powder diffraction*

Since the Fe atoms are arranged in an *F*-centered cell and the *F*-centering is only broken by weakly diffracting nitrogen atoms, the h00 reflections with even *h* (allowed for *F*-centering) are much stronger than the reflections with odd *h* (forbidden for *F*-centering). Within the diffraction range of CuKα radiation, the strong 200 and 400 and weak 100 reflections are visible.

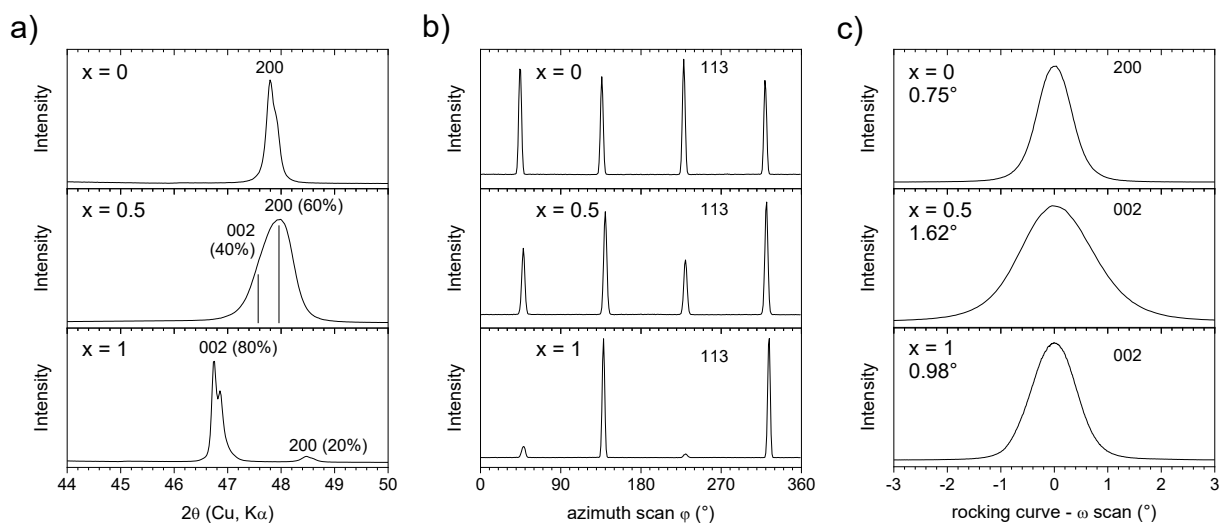

Fig.S13. X-ray diffraction of $Fe_{4-x}Ge_xN$ thin films on MgO substrate. (a) Diffraction pattern of *200/002* reflection, (b) azimuth scan of *113* reflection (*hkl* indexes of the primitive cell), (c): Rocking-curve scan of *200/002* reflection.

*Indentation modulus and hardness*

The results of indentation experiments show the same elastic Young modulus for *x* = 0 and 0.5 and a significant decrease in the modulus for *x* = 1. On the other hand, the hardness determined by this method exhibits an increase from *x* = 0 to *x* = 0.5 and then a slight decrease, but with *x* = 1 remaining harder than *x* = 0, see Table S1. For the confrontation of experimental values with calculation, we have performed the DFT electronic structure calculation of elastic properties. The parameters $C_{11}$, $C_{12}$ and $C_{44}$ were evaluated for cubic structure of *x* = 0, and parameters $C_{11}$, $C_{33}$, $C_{12}$, $C_{13}$, $C_{44}$ and $C_{66}$ for the tetragonal structure of *x* = 1, where we supposed full occupation of the Fe2-site by Ge and the Fe1-site by Fe, as confirmed by Mössbauer results. Using these parameters, several elastic properties were calculated, including elastic wave velocity and Debye temperature, see Table S2. Hardness was calculated using models suggested in [50]. Generally, the calculated elastic quantities increased for *x* = 1. This trend could be expected, since Ge introduces more covalent character into the structure, so it increases the strength of the bonds [51]. The calculated hardness is in quite good agreement with experiment as regards the trend as well as the absolute values. They are also in good agreement with previous literature data of $Fe_4N$ being consistently around 8 GPa [52,53].

On the other hand, the calculated elastic moduli show an opposite trend to the experiment. Possible explanations may be related to the lattice mismatch between the film and substrate, which increases with *x*, and the resulting defects in the thin film. Previously obtained experimental data on Young modulus of $Fe_4N$ are also rather inconsistent, ranging from 160 to 210 GPa [52,53,54,55]. These differences in experimental data may be caused by different types of samples (thin layers, bulk samples, etc.), their microstructure and different methods used. The range of previous literature data thus includes both our experimental and calculated values for the elastic modulus of $Fe_4N$. Comparison with literature data therefore does not unambiguously distinguish which of these values is more accurate. Regarding $Fe_3GeN$, no literature data on elastic properties are available.

Table S1. Average values of Young modulus (*E*) and hardness (*H*) determined by indentation and by DFT calculation.

|  | $E_{exp}$ (GPa) | $H_{exp}$ (GPa) | $E_{DFT}$ (GPa) | $H_{DFT}$ (GPa) |
|---|---|---|---|---|
| $Fe_4N$ | 203 ± 8 | 8.1 ± 0.2 | 162 | 7.6 |
| $Fe_{3.5}Ge_{0.5}N$ | 204 ± 5 | 9.5 ± 0.4 |  |  |
| $Fe_3GeN$ | 156 ± 4 | 8.8 ± 0.4 | 222 | 9.2 |

Table S2. Results of DFT calculation (Wien2k package) of elastic properties (IRelast package).

| Mechanical properties | $Fe_4N$ | $Fe_3GeN$ |
|---|---|---|
| Bulk modulus, *K* (GPa) | 193.91 | 227.14 |
| Young modulus, *E* (GPa) | 161.74 | 221.53 |
| Shear modulus, *G* (GPa) | 59.42 | 82.82 |
| Poisson's coefficient, *n* | 0.36 | 0.34 |
| Hardness, *H* (GPa) | 7.6 | 9.2 |
| Zener elastic-anisotropy ratio, *A* | 0.45 | 0.57 |
| Pugh ratio, *k* | 0.31 | 0.36 |
| Transverse elastic wave velocity, $v_t$ (m/s) | 2861.85 | 3205.13 |
| Longitudinal elastic wave velocity, $v_l$ (m/s) | 6135.70 | 6470.94 |
| The average elastic wave velocity, $v_m$ (m/s) | 3222.40 | 3597.53 |
| Debye Temperature, $\theta$ (K) | 433.14 | 489.57 |
| Stiffness tensor |  |  |
| *C11* (GPa) | 320.25 | 388.49 |
| *C33* (GPa) |  | 423.53 |
| *C12* (GPa) | 130.74 | 148.77 |
| *C13* (GPa) |  | 136.71 |
| *C44* (GPa) | 43.08 | 59.99 |
| *C66* (GPa) |  | 62.79 |

## *DFT calculations*

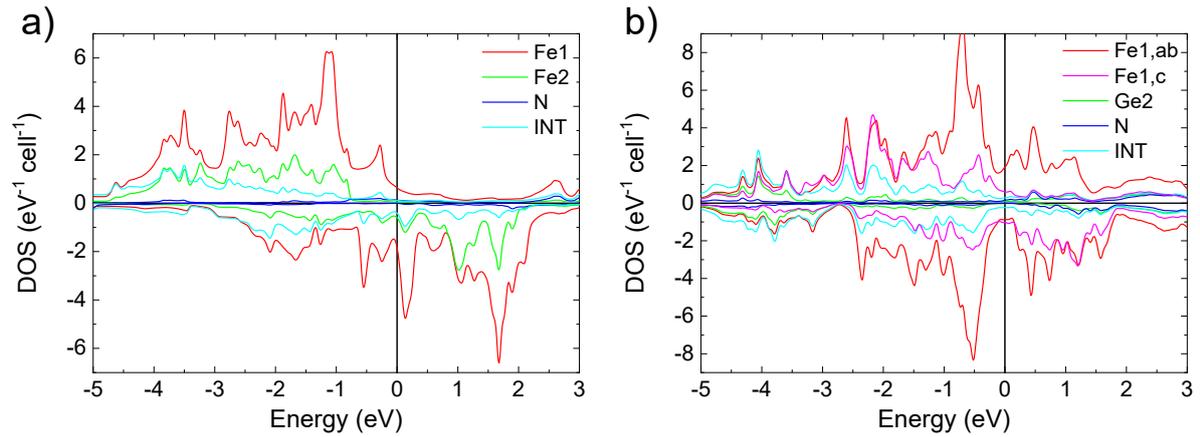

Fig. S2. DOS for (a) *x* = 0 and (b) *x* = 1. Partial DOS for individual atoms and for interstitial region (INT), which cannot be unambiguously assigned to a specific atom, are shown. Note the reduction of spin polarization for *x* = 1.

## *Hall and Nernst effect*

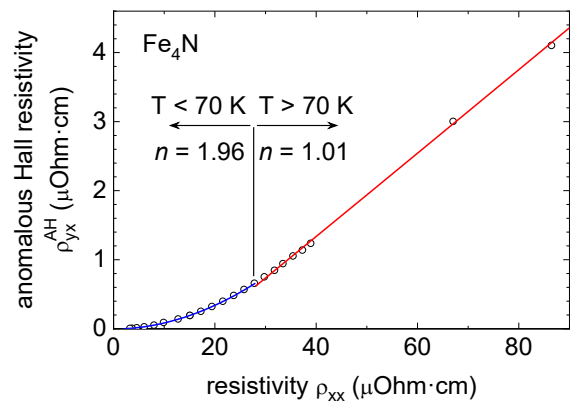

Fig. S3. The fit of the relation $\rho_{yx}^{AH} \sim \lambda \rho_{xx}^{n}$ for Fe$_4$N.

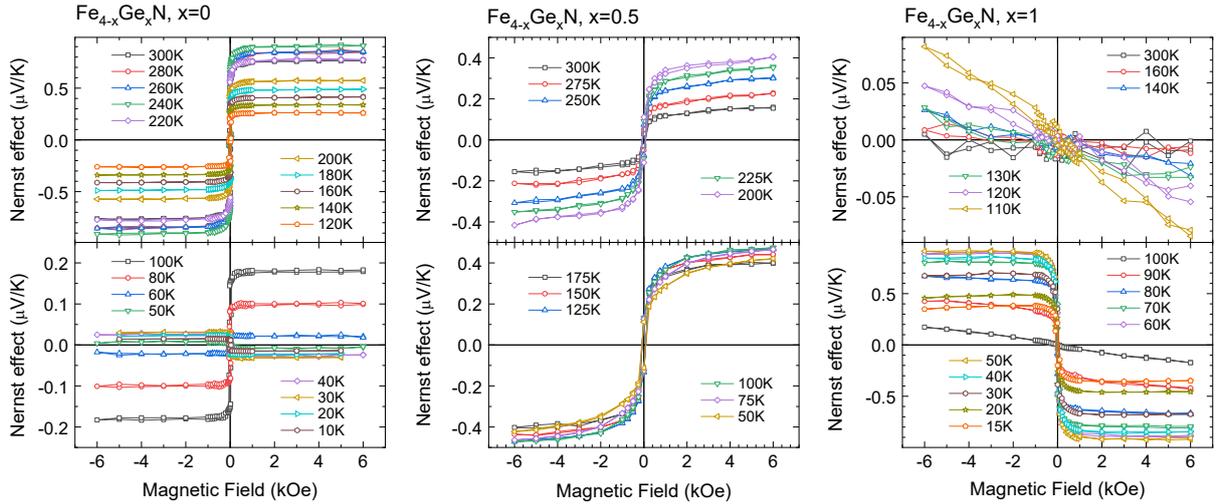

Fig. S4. Nernst effect hysteresis loops for $Fe_{4-x}Ge_xN$.

## $^{57}Fe$ conversion-electron Mössbauer spectroscopy (CEMS)

In contrast to the traditional transmission Mössbauer spectroscopy, conversion-electron Mössbauer spectroscopy (CEMS) uses the conversion electrons emitted from the sample after the nuclear resonant absorption of the γ photons, with a limited information depth of roughly 200 nm. The CEMS spectra of the $Fe_{4-x}Ge_xN$ samples with $x$ = 0.5 and 1.0 acquired at room temperature are shown in Fig. S5, the hyperfine parameters of the fitted components are provided in Table S3. The spectrum of $Fe_3GeN$ at room temperature contains only three doublets D1, D2 and D3, indicating a fully paramagnetic state in accordance with $T_C$ ≈ 100 K evidenced by temperature-dependent magnetization shown in Fig. 6 in the main text. From the comparison of the experimental isomer shift (IS) with DFT calculation results (see Table S4), we can conclude that Ge occupies the *4b* site, related to the large site in the cubic antiperovskite structure of γ'-$Fe_4N$. Then the two dominating doublets D1 and D2 in the 1:2 ratio correspond to the Fe1c and Fe1ab sites, respectively, forming an elongated octahedron with a nitrogen atom in the center. The higher *IS* than in the γ'-$Fe_4N$ phase, *IS* (Fe1 in $Fe_4N$) = 0.30 mm/s [56], reflects the decrease in the occupation of (conduction) s-electrons. Fe1c and Fe1ab have remarkably different electronic properties, *e.g.* they exhibit dominating local and itinerant magnetism in the ground state at low temperatures, respectively [57]. The related variations in electron density are correlated with observed differences in *IS*. Nitrogen vacancies introduce different coordination environments of Fe atoms, resulting in a large number of components in the spectrum with various *IS* and quadrupole splitting (*QS*) values. However, considering the spectrum's shape, these components would exhibit strong correlations when fitted. Finally, a minor doublet D3 can be ascribed to Fe diffused in the Al coating layer.

The spectrum of the $Fe_{3.5}Ge_{0.5}N$ sample still contains a magnetically ordered component, whose magnetic moment (and thereby hyperfine magnetic field at Fe nuclei) fluctuates at the timescale comparable to the characteristic time window of the method, $\tau \sim 10^{-7}$ s. As we can assume a random distribution of Ge among the *4b* site in the $I4/mcm$ structure, obtaining an unambiguous fit remains challenging mainly due to the significant variation in Fe coordination environments, including Ge atoms and potential nitrogen vacancies [58]. Therefore, the magnetic component was approximated by a single sextet (S) within a Blume-Tjon relaxation model, in which the hyperfine magnetic field oscillates between two antiparallel orientations along the EFG principal axis [48]. This component encompasses the signal from Fe with fewer Ge neighbors (both the Fe in *4b* site and their neighboring Fe1) and has a lower isomer shift similar to the undoped γ'-$Fe_4N$ [56]. The fit provided an average frequency of hyperfine-field fluctuations of $1.3(2) \cdot 10^7$ s$^{-1}$. Fe atoms with more Ge neighbors exhibit more rapid magnetic relaxation and manifest as paramagnetic doublets in the spectrum. Given the local similarity of Ge-rich areas to Fe$_3$GeN, two doublets D1 and D2 in a 1:2 ratio were used to model Fe11- and Fe12-like species, respectively. A higher quadrupole splitting than in Fe$_3$GeN relates to the reduced symmetry of Fe sites, while a lower average isomer shift is consistent with a higher density of conducting electrons (see Fig. S6 for a direct comparison of the spectra). For comparison, we have calculated some of the parameters for Fe$_3$GeN using DFT method according to procedures applied in [59]. We have considered structural models with Ge in all the Fe1c, Fe1ab and Fe2 sites. A comparison of the experimental and calculated IS clearly confirmed that Ge occupies the Fe2 site.

Table S3. Hyperfine parameters from the fit of room-temperature $^{57}$Fe CEMS spectra of the Fe$_{4-x}$Ge$_x$N ($x$ = 0.5 and 1) films. $\Delta QS$ represents the width of the $QS$ distribution, $B_{hf}$ the amplitude of the fluctuating hyperfine field, $A$ the intensity of the components and $\Gamma_{FWHM}$ the linewidth. Comparison with DFT calculation.

| Sample | Comp. | Site | IS (mm/s) | QS (mm/s) | ΔQS (mm/s) | $B_{hf}$ (T) | $\Gamma_{FWHM}$ (mm/s) | A (%) |
|---|---|---|---|---|---|---|---|---|
| Fe$_3$GeN | D1 | Fe1c | 0.50(2) | 0.40(4) | 0.08(3) | - | 0.29† | 31(1)‡ |
| | D2 | Fe1ab | 0.41(1) | 0.17(2) | 0.04(5) | - | 0.29† | 62(1)‡ |
| | D3 | Al(Fe) | 0.13(1) | 0.39(3) | 0.00 | - | 0.29† | 7(1) |
| Fe$_{3.5}$Ge$_{0.5}$N | S | | 0.31(5) | -0.01(5) | - | 16.5(8) | 0.27* | 48(2) |
| | D1 | | 0.42(2) | 0.40(2) | - | - | 0.27(2) | 17(1)⁕ |
| | D2 | | 0.38(2) | 0.52(2) | - | - | 0.43(3) | 35(1)⁕ |

*fixed, †,‡,⁕ bound parameters

Table S4. Hyperfine parameters of the Fe$_3$GeN from the DFT calculation. IS represents isomer shift, QS quadrupole splitting, B$_{hf}$ the amplitude of the fluctuating hyperfine field, and A the intensity of the components. Structural parameters were optimized by DFT.

| Ge site | Fe Site | IS (mm/s) | QS (mm/s) | $B_{hf}$ (T) | A (%) |
|---|---|---|---|---|---|
| Fe2 | Fe1c | 0.52 | 0.37 | 13.4 | 33 |
| Fe2 | Fe1ab | 0.40 | 0.41 | 4.7 | 66 |
| Fe1c | Fe2 | 0.27 | 0.33 | 19.1 | 33 |
| Fe1c | Fe1ab | 0.33 | 0.99 | 22.9 | 66 |
| Fe2* | Fe1c* | 0.53* | 0.15* | 8.8* | 33* |
| Fe2* | Fe1ab* | 0.45* | 0.18* | 4.2* | 66* |

* Structure from Scholz *et al.* [19]

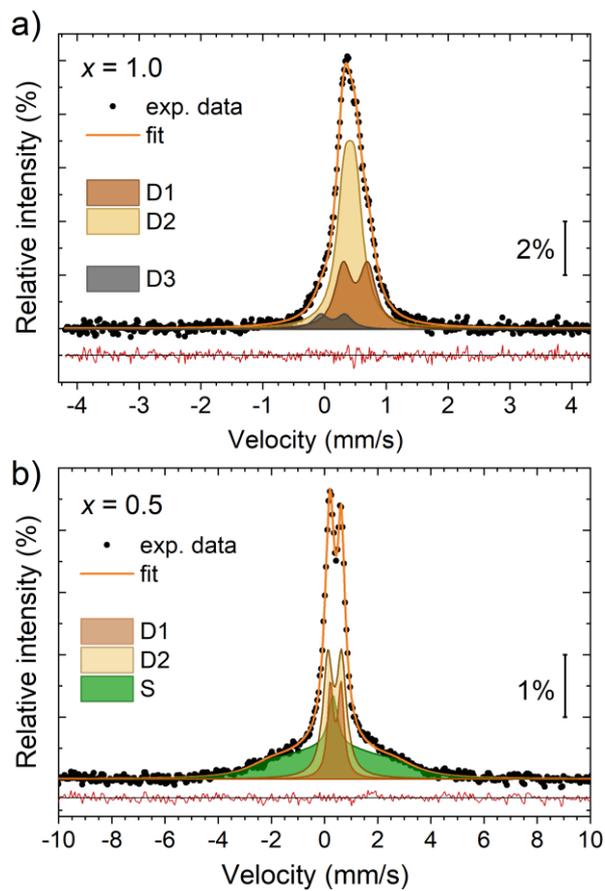
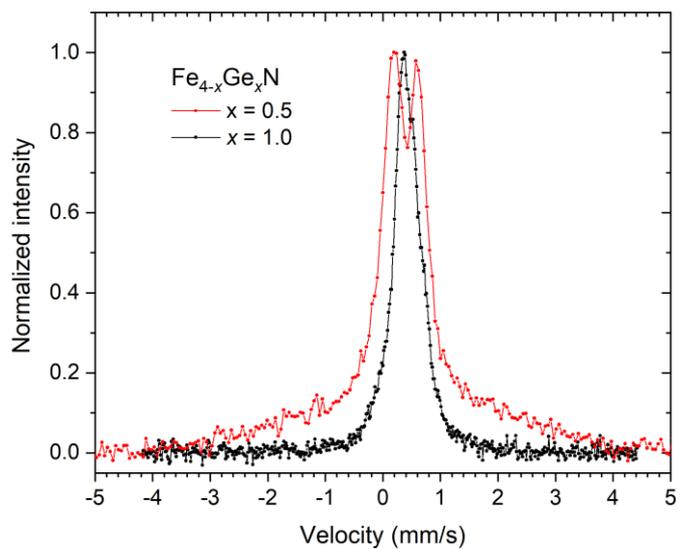

Fig. S5. Room-temperature $^{57}$Fe CEMS spectra of the Fe$_{4-x}$Ge$_x$N films, the red line shows the difference between the data and the fit (note the different velocity scales).

Fig. S6. A comparison of room-temperature $^{57}$Fe CEMS spectra of the Fe$_{4-x}$Ge$_x$N (x = 0.5, 1.0) films on the MgO substrate, each spectrum was normalized to the maximum intensity.